\documentclass[11pt]{article}
\usepackage{blois,epsfig}

\def\be{\begin{equation}}
\def\ee{\end{equation}}
\def\bea{\begin{eqnarray}}
\def\eea{\end{eqnarray}}

\newcommand{\ba}{\begin{eqnarray}}
\newcommand{\ea}{\end{eqnarray}}
\newcommand{\bi}{\begin{itemize}}
\newcommand{\ei}{\end{itemize}}

\newcommand{\fig}{Fig.~}

\newcommand{\la}{\label}

\begin{document}
\vspace*{2cm}
\begin{center}
\Large{\textbf{XIth International Conference on\\ Elastic and Diffractive Scattering\\ Ch\^{a}teau de Blois, France, May 15 - 20, 2005}}
\end{center}

\vspace*{2cm}
\title{GLUEBALLS AND THE POMERON -- A LATTICE STUDY\footnote{Work done
in collaboration with Michael~Teper.}}

\author{ HARVEY~B.~MEYER }

\address{Deutsches Elektronen-Synchrotron DESY,\\
Platanenallee 6, D-15738 Zeuthen}

\maketitle\abstracts{
We perform lattice calculations of the lightest
$J=0,2,4,6$ glueball masses in the D=3+1 SU(3) gauge
theory and extrapolate to the continuum limit.
Assuming that these masses lie on linear Regge
trajectories we find a leading glueball trajectory
$\alpha(t)=0.93(24) + 0.28(2)\alpha^\prime_R t$, where
$\alpha^\prime_R \simeq 0.9 \, \mathrm{GeV}^{-2}$ is the
slope of the usual mesonic
Regge trajectories. This glueball trajectory has an intercept
and slope similar to that of the Pomeron.
We contrast this with the situation
in D=2+1 where the leading glueball Regge trajectory
is found to have too small an intercept to be important
for high-energy cross-sections.
We attempt to interpret the observed states and trajectories in
terms of open and closed string models of glueballs.
}

\section{Introduction and main results}
\label{section_intro}
The Pomeron trajectory is qualitatively different from other 
Regge trajectories in that it is
much flatter ($\alpha^{\prime}$ much smaller)
and has a higher intercept\cite{book},
$\alpha_P(t=m^2)\simeq 1.08 + 0.25 m^2/{\rm GeV}^2$. 
There has been a long-standing speculation that
the physical particles on the trajectory might be glueballs. 
If we consider the high-energy hadron scattering 
in a world deprived of the $u$, $d$ and $s$ quarks,
it is difficult to imagine that the total cross-sections should
behave very differently from those in the real world.
For instance, in leading-logarithmic perturbative
calculations\cite{book}, only the gluonic field contributes to the Pomeron.
Thus it is reasonable to expect that the Pomeron phenomenon would also 
be observed in the absence of light quarks\cite{peschanski}.
This constitutes the main motivation for the present investigation: 
we use numerical lattice techniques to 
investigate whether the mass spectrum of the SU(3) pure gauge theory 
is consistent with approximately straight Regge trajectories,
the leading one of which possesses the properties of the 
phenomenological Pomeron.

Our lattice calculations employ the standard Wilson action on a cubic 
lattice. We calculate ground and excited state masses, $m$, from 
Euclidean correlation 
functions using standard variational techniques. We calculate the
string tension, $\sigma$, by calculating the mass of a flux loop 
that closes around a spatial torus. We perform calculations
with a 2-level algorithm\cite{2leva} at lattice 
spacings $a \simeq 0.10 - 0.05 \, \mathrm{fm}$. The calculations are on
lattices ranging from $16^3 36$ to $32^3 48$, corresponding to
a spatial extent $L \simeq 1.5 \, \mathrm{fm}$. 
We also perform a calculation on a lattice of size $2 \, \mathrm{fm}$ so as
to check that any finite volume corrections are small. We
extrapolate the calculated values of the
dimensionless ratio $m/\surd\sigma$ to
$a=0$ using an $a^2\sigma$ correction. 

On the lattice, the problem with the identification of
the lightest $J\geq 4$ states is that its cubic 
symmetry group is much smaller than the continuum
rotation group and has just a few irreducible representations. 
Nonetheless, as $a\to 0$ an energy 
eigenstate belonging to one of these lattice representations
will tend to some state that is labelled by spin $J$. So using 
continuity we can refer to a state at finite $a$ as being of `spin
$J$' if $a$ is small enough. At $a=0$ 
a state of spin $J$ is a multiplet of $2J+1$ degenerate 
polarisations; if we now increase $a$ from zero, these  $2J+1$ polarisations
will in general appear in different lattice representations, and the 
degeneracy will be broken at $O(a^2)$. So a particular polarisation of the ground state of 
spin $J=4,5, 6, ...$ will in general be a (highly) excited state in some lattice
representation, thus complicating its identification.
If we can perform this identification, then we can extrapolate 
the mass of the state to $a=0$, so obtaining the mass of
the lightest state of spin $J$. Our identification technique\cite{hspin}
is to perform a Fourier analysis
of the rotational properties of any given lattice eigenstate.
For this we may use as a `probe' a set of fuzzy Wilson loops 
based on the propagator within one time-slice of a massive scalar field
in the gauge field background.
If we keep its mass fixed in physical units, it is guaranteed 
that the rotational invariance of the propagator is restored as $a\to0$ 
(upon averaging over the gauge field). 
The Wilson loops, which have typically a size of 0.5fm, then have a
rotational symmetry that is broken only by $O(a^2)$ effects,
so that we can probe the rotational properties of the glueballs 
under rotations finer than $\pi/2$ to that accuracy. Subsequently we found more heuristic
techniques to construct the probe operators, which however are computationally
much cheaper and provided as good rotational properties as the propagator construction
at the lattice spacings we were working at.
\begin{figure*}[t]
\centerline{
\psfig{file=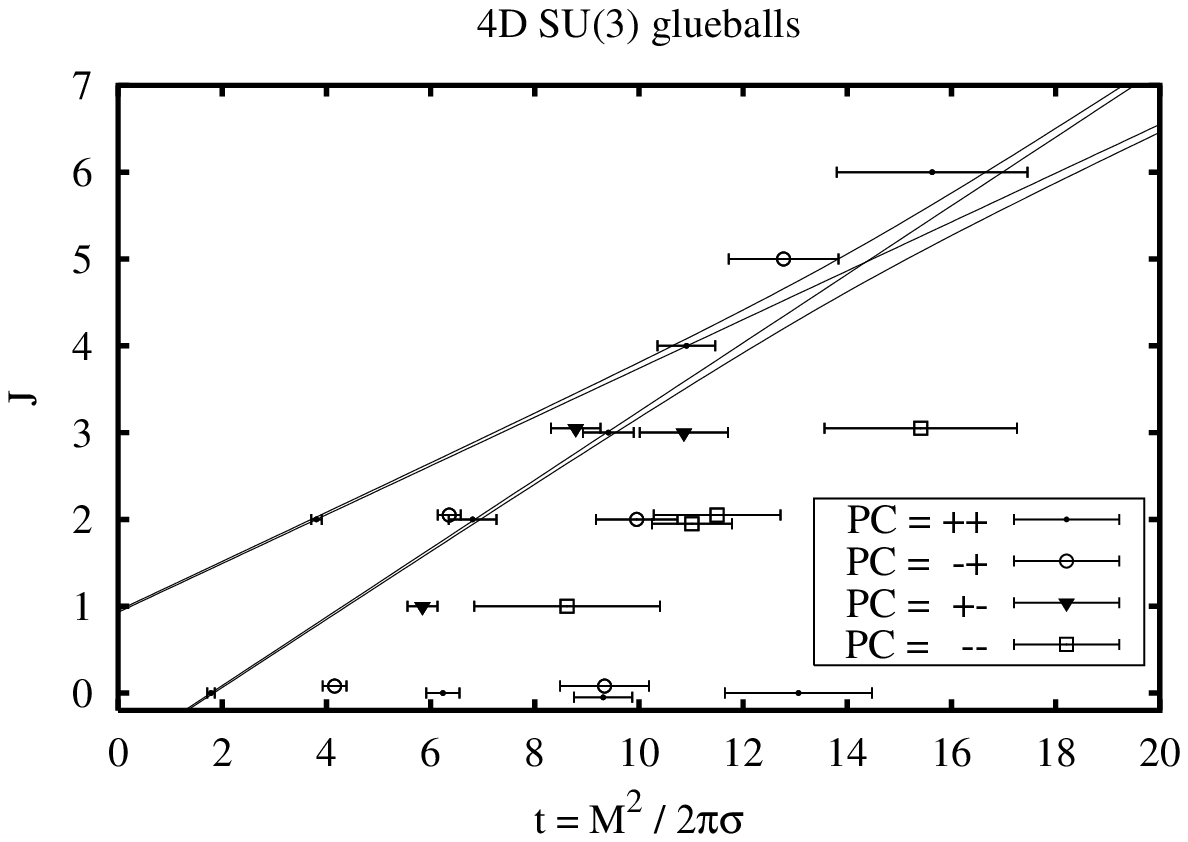,angle=0,width=7.5cm}\\	
\psfig{file=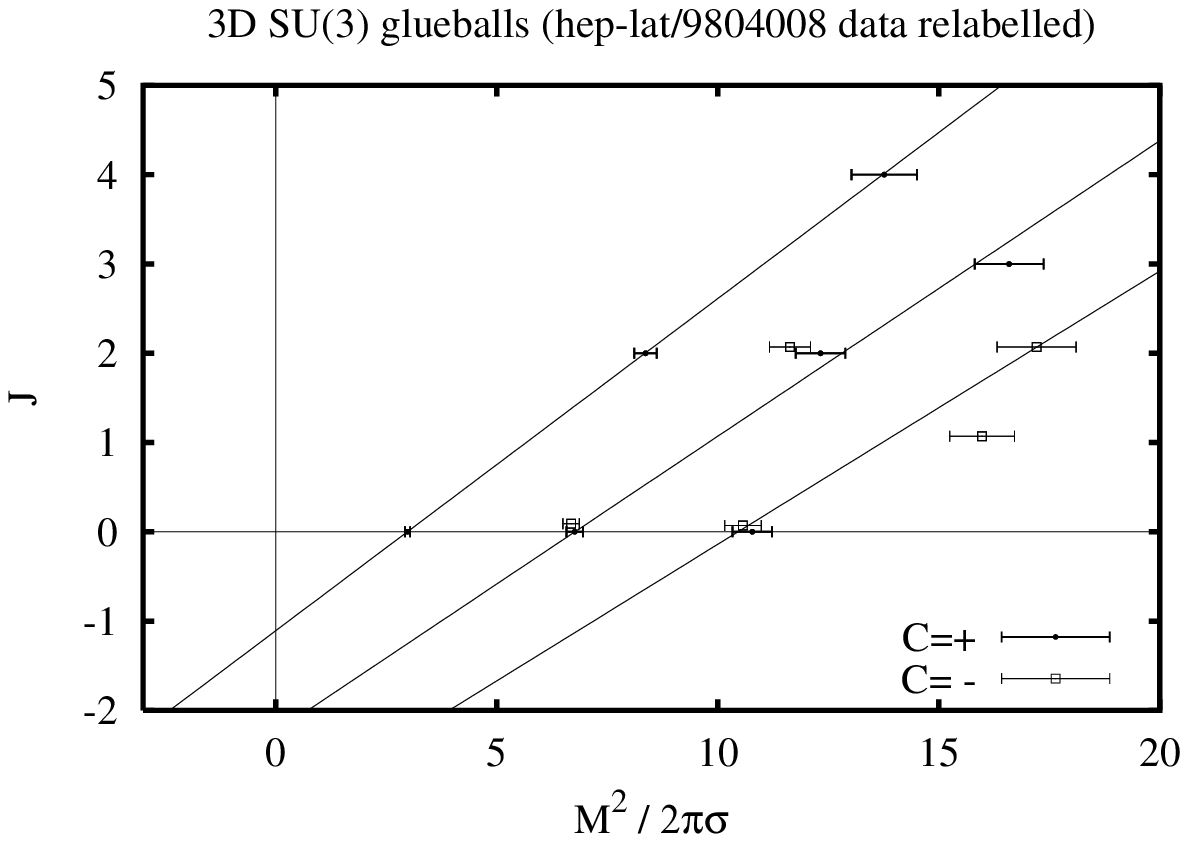,angle=0,width=7.5cm} }
\caption{Chew-Frautschi plot of the continuum $SU(3)$ gauge theory, 
in $D=3+1$ (left) and $D=2+1$ (right).}
\la{fig:cf}
\end{figure*}

Having extrapolated our glueball masses to the continuum limit,
we plot the squared masses against the spins in a Chew-Frautschi 
plot\cite{hspin}, as in \fig\ref{fig:cf} (left). 
We now assume that the states fall on approximately linear Regge trajectories. 
In that case the leading trajectory clearly passes through the lightest
$J=2$ and $J=4$ glueballs. We note that there is no odd $J$ 
state on this trajectory: it is even signature.
The parameters of the trajectory are 
$2\pi\sigma\alpha'=0.281(22),~\alpha_0=0.93(24),$
in agreement with the phenomenological Pomeron
if we recall that the usual mesonic trajectories have slopes
$\alpha'_{\cal R} \simeq \frac{1}{2\pi\sigma}\simeq 0.9 \, \mathrm{GeV}^{-2}$.
Of course, in comparing our leading pure-glue trajectory with the 
phenomenological Pomeron we should not ignore the fact that the
latter will mix with the flavourless mesonic trajectory: 
this effect will presumably increase the intercept and the slope of the Pomeron. 

We can identify  the sub-leading glueball trajectory in \fig\ref{fig:cf} as well.
It contains the lightest $J=0$ glueball, the first-excited $J=2$ glueball
and the lightest $J=3$ glueball. In striking
contrast to what one finds for the usual mesonic trajectories,
this secondary trajectory is clearly not parallel to the leading one.
As we shall see, this is not unexpected in a string picture of glueballs. 
The trajectories, if linear, would cross somewhere near $J=5$; 
because of unitarity the true trajectories will not
cross but rather repel, as suggested on the figure.

The right plot shows the spectrum\cite{teper98} in $D=2+1$\footnote{We do not refer to parity, 
because in two space dimensions one has automatic parity-doubling for $J\not= 0$.}.
In contrast to $D=3+1$, a linear trajectory between the 
lightest $J=2$ and $J=4$ states passes through the lightest
$J=0$ state. Between them the $J=0,2,4$
states provide strong evidence for the approximate linearity
of the trajectory. In contrast to $D=3+1$ the secondary
trajectory is approximately parallel to the leading one. 
The parameters of the leading trajectory are
$ 2\pi\sigma\alpha'=0.384(16),~\alpha_0=-1.144(71)$.
Unlike the intercept, the slope of the trajectory
is similar to what we found in $D=3+1$.

\section{Interpreting the glueball spectrum in terms of string models}
\label{section_strings}
A natural model for a high $J$ meson is to see it as
a rotating string with a $q$ and $\bar{q}$ at its ends
and at the classical level this leads to linear 
Regge trajectories, $J \stackrel{J\to\infty}{\sim} \frac{1}{2\pi\sigma} m^2$.
If we now go to the pure gauge theory, this simple `open string'
model has an immediate analogue: two gluons joined
by a string containing flux in the adjoint representation\cite{kaidalov}.
The rotating adjoint string  produces a linear Regge trajectory
$J \stackrel{J\to\infty}{\sim} \frac{1}{2\pi\sigma_A} m^2 
\simeq \frac{2}{9\pi\sigma} m^2 $ (assuming Casimir scaling). The trajectory
is  much flatter than the usual mesonic Regge trajectory, although not quite
as flat as the phenomenological Pomeron or the leading glueball
trajectory that we identified. 
Since the adjoint string comes back to itself under $C$, $P$ or
rotations of $\pi$, its spectrum contains 
$0^{++}, 2^{++}, 4^{++}, ...$ states, as expected for an even-signature Pomeron.

An equally natural model\cite{isgur} pictures glueballs as composed
of a closed loop of fundamental flux with no constituent gluons
at all (one might expect some glueball states to be open strings
and others to be closed strings, with mixing between the two).
There are phonon-like excitations of this closed string
which propagate around it and
contribute to both its energy and its angular momentum.
The whole loop can rotate around its diameter, obtaining angular momentum
that way too. If one considers the set of states where
the angular momentum is purely phononic one obtains an
asymptotically linear Regge trajectory with slope\cite{thesis}
$J \stackrel{J\to\infty}{\sim} \frac{1}{8\pi\sigma} m^2 $
while for a loop with purely (non-relativistic) orbital motion one
obtains a linear trajectory with
$ J \stackrel{J\to\infty}{\sim} \frac{3\surd 3}{32\pi\sigma} m^2 $.
In either case the slope 
$\alpha^\prime \simeq 0.2 - 0.3 \, \mathrm{GeV}^{-2}$ is in
the right range for the Pomeron.
The orbital trajectory leads to a trajectory of states with positive parity and
$C~=~(-1)^{J}$, $J=0,1,2\dots$
For the leading phononic trajectory, the most striking feature is the 
absence of a $J=1$ state: apart from a $0^{++}$ state, 
all $PC$ combinations are then expected for $J\geq2$.

The SU(3) gauge theory in $D=2+1$ is linearly confining
and therefore an effective string theory description is equally well 
motivated. Since the rotating open string lies 
in a plane, it provides a natural model for glueballs in
two space dimensions. It will contribute states with $J$ even and $C=+$. 
The closed string is also a possibility; the quantum numbers for the leading $C=+$ 
and $C=-$ phononic trajectories are
$0^{++},~2^{\pm+},~3^{\pm+},~4^{\pm+}\dots$ and
$0^{--},~2^{\pm-},~3^{\pm-},~4^{\pm-}\dots $
In the simplest form of the model, the two trajectories are degenerate.

Returning to the data, the $D=2+1$ leading trajectory contains only even $J$ states
with $C=+$ and so is naturally interpreted as arising
from a rotating open string. Since the
intercept is sufficiently  low, it can and does include a 
$J=0$ state, in contrast to the case of 3 spatial dimensions.
The first subleading trajectory has no $J=1$ state, although
it contains a $J=3$ state, and possesses a $C=+/-$ degeneracy
for the lower $J$ where we have reliable calculations.
All this strongly suggests a phononic trajectory of the closed string.

In $D=3+1$, for $J\leq 4$ the leading trajectory contains only even spin
states with $PC=++$. This again suggests that the trajectory arises
from a rotating open string carrying adjoint flux between the
gluons at the end points. The subleading trajectory has no $J=1$ state although it does
appear to have a $J=3$ state, again a feature
of the closed string phononic spectrum.

The states one might expect to lie along the \emph{odderon}\cite{odderon} 
are the lightest odd $J$, $PC=--$ glueballs. From \fig\ref{fig:cf}
we see that a trajectory defined by the lightest $1^{--}$ and 
$3^{--}$ will have a slope similar to the Pomeron and a very low,
negative intercept. However, if the leading trajectory 
has an intercept around unity, as claimed phenomenologically,
then the lightest $1^{--}$ glueball cannot lie on it, but will
rather lie on a subleading trajectory. To test this possibility
we need a good calculation of the lightest $5^{--}$ glueball,
something we do not yet have. 
\section{Conclusions}
\label{section_conclude}
Using novel lattice techniques, we have calculated the masses of 
higher spin glueballs in the continuum limit of the SU(3) gauge
theory. We find a leading $PC=++$ glueball trajectory 
$\alpha_P(t) = 0.93(24) + 0.25(2) t/{\rm GeV}^2$ (assuming linearity)
which is entirely consistent with the phenomenological Pomeron. 
The sub-leading trajectory has a larger slope and eventually 
`crosses' the Pomeron. We argue that such a rich Regge structure 
for glueballs occurs naturally within string models. 

In contrast to this, we find that in 2+1 dimensions the intercept
of the leading trajectory is negative so that it
does not contribute significantly to scattering at 
high energies. We find evidence that the leading trajectory
is an open string while the non-leading one is a closed string.
In this case we have enough accurately calculated glueball states 
along the leading trajectory to demonstrate its approximate linearity.

In a world deprived of the $u$, $d$ and $s$ quarks, the mass gap
would be given by the lightest glueball, $m_G\simeq1.6$GeV;
the Froissart bound is then stronger by two orders of magnitude\cite{nicolescu}.
Experimentally, the high-energy $pp$ cross-section lies only slightly
below that bound. If the  cross-section is found to exceed it at the LHC,
then it will definitely
be necessary to include the effects of light quarks in the description
of the hadronic wave-functions at that energy.


\end{document}